\documentstyle[11pt]{article}\textheight 230mm\textwidth 150mm
            \newcommand{\be}{\begin{eqnarray}}
            \newcommand{\ee}{\end{eqnarray}}
\newcommand{\e}[1]{\label{e:#1}\end{eqnarray}}
     \newcommand{\eg}{{\em e.g.\ }}
            \newcommand{\ie}{{\em i.e.\ }}
            \newcommand{\ga}{{\gamma}}
 \newcommand{\Ga}{{\Gamma}}
  \newcommand{\baGa}{{\bar{\Gamma}}}
            \newcommand{\la}{{\lambda}}
            
            \newcommand{\del}{{\delta}}
           \newcommand{\ra}{{\rightarrow}}
 \newcommand{\lea}{{\leftarrow}}
            
            \newcommand{\Lra}{{\Leftrightarrow}}

            \newcommand{\beq}{\begin{quote}}
            \newcommand{\eq}{\end{quote}}
            
            \newcommand{\al}{\alpha}
            \newcommand{\half}{\frac{1}{2}}
            \newcommand{\ben}{\begin{enumerate}}
            \newcommand{\een}{\end{enumerate}}
\newcommand{\bea}{\begin{array}}
            \newcommand{\ea}{\end{array}}
            \newcommand{\bit}{\begin{itemize}}
            \newcommand{\ei}{\end{itemize}}

            \newcommand{\nn}{\nonumber}
            \newcommand{\r}[1]{(\ref{e:#1})}
            \newcommand{\edfl}[1]{\label{#1}\end{df}}
             \newcommand{\ex}{{_{\rm ext}}}
\newcommand{\ep}{{\varepsilon}}
\newcommand{\bd}{{\stackrel{\lea}{\partial}}}
\newcommand{\bad}{{\bar{\partial}}}
\newcommand{\bbad}{{\stackrel{\lea}{\bar{\partial}}}}
\newcommand{\baC}{{\bar{\cC}}}
\def\fa{field-antifield }
\def\R{\right}
\def\L{\left}
\def\tilde{\widetilde}
\def\bar{\overline}
\def\d{{\partial}}

\def\cA{{\cal A}}

\def\cC{{\cal C}}
\def\cD{{\cal D}}

\def\cM{{\cal M}}

\def\IJMPA{{\sl Int. J. Mod. Phys. }A}
\def\MPLA{{\sl Mod. Phys. Lett.} A}

\begin{document}
            \begin{titlepage}
            \newpage
            \noindent
            G\"oteborg ITP 97-03\\
            April 1997\\
												hep-th/9704209\\
(Appendix B inserted May 1997 and a new introduction August 1997)
            \vspace*{35 mm}
            \begin{center}
{\LARGE\bf Antisymplectic gauge theories}
\end{center}
            \vspace*{15 mm}
        \begin{center}{\large \bf Igor Batalin}
\footnote{On leave of absence from
P.N.Lebedev Physical Institute, 117924  Moscow, Russia\\ \hspace{3.5mm}
 E-mail:
batalin@lpi.ac.ru } and {\large \bf Robert Marnelius}
\footnote{E-mail:
tferm@fy.chalmers.se} \\
          \vspace*{10 mm} {\sl
            Institute of Theoretical Physics\\ Chalmers University of
            Technology\\ S-412 96  G\"{o}teborg, Sweden}\end{center}
            \vspace*{27 mm}
            \begin{abstract}
A general field-antifield BV formalism for antisymplectic
 first class constraints is 
proposed. It is as general as the corresponding symplectic BFV-BRST
 formulation and it
is demonstrated to be consistent with a previously proposed  formalism for
antisymplectic second class constraints through a  generalized 
conversion to
corresponding first class constraints. Thereby the basic concept
 of gauge symmetry is
extended to apply to quite a new class of gauge theories
 potentially possible to
exist. 
  \end{abstract} \end{titlepage}
            \newpage
            \setcounter{page}{1}
            \section{Introduction}
The field-antifield BV formalism \cite{BV} is a 
 Lagrangian path integral method to quantize
general gauge theories (important early contributions are 
\cite{ZJ}-\cite{WH}). It has been shown to work for
 an ever increasing number of models. In the
BV formalism one introduces antifields  with opposite 
Grassmann parities to all field and ghost
variables. It involves in a crucial way also an 
antibracket operation and a nilpotent differential
$\Delta$-operator. The understanding of the 
formalism was further deepened in
\cite{Vol}-\cite{Schw} and in \cite{BT93}-\cite{Trip}. 
In the  approach of the latter papers a
coordinate invariant general covariant formulation 
was developed. The field-antifield variables are
here considered as arbitrary coordinates on an 
antisymplectic manifold $\cM$. (The standard BV
formalism may then be viewed as  formulated in 
terms of antisymplectic Darboux coordinates.) In
this formalism the geometric coordinate invariant 
properties is  formally demonstrated.
Furthermore, the formalism specifies the conditions 
a consistent invariant measure density has to
satisfy. It involves also new ingredients like a hypergauge formulation
\cite{BT93,Trip} and a multilevel formalism
\cite{BT93-2}. Gauge invariance is demonstrated in 
general terms. Among the further generalizations
are deformed $\Delta$-operators in \cite{BT94-1}
 and an Sp(2) version in \cite{Trip,GTrip}.
 In \cite{BT93} it was also shown
 how antisymplectic second class constraints may be introduced and 
treated consistently within this  formalism.  
In this paper we continue this set of formal
generalizations with still another one.  Here we 
show that the path integral may be formulated 
on a large antisymplectic manifold also in the 
presence of antisymplectic first class constraints.
All required conditions are shown to be formally 
satisfied. This is therefore a major
further generalization of the general covariant BV 
formalism. The beautiful general mathematical
structure of the BV formalism is thereby further extended. 
However, it remains to demonstrate the
existence of examples which satisfy the generalizations
 suggested by the obtained formal results.
Although we expect them to exist this is certainly a
 nontrivial issue. Anyway the formal results
suggest alternative formulations which could turn 
out to be useful. Particularly the results of the
present paper could allow for formulations with 
specific  global symmetries which are  preferable
for some reasons. At a more speculative level 
our results show the existence of new types of gauge
theories in an antisymplectic quantum theory in the spirit of
\cite{Vol}-\cite{Khud}.

In section 2 we recapitulate some basic properties of 
the general covariant BV formalism. In section
3 and appendices A and B we present our formulation 
and its formal properties. In section 4 and
appendix C we  consider then a generalized conversion  
of antisymplectic second class constraints
into
 corresponding first class
ones by means of an  extension of the
\fa manifold
$\cM$. This provides for an explicit formal verification 
of the formalism. Throughout the paper we
make use of deWitt's compact notation which reduces 
the treatment to a finite dimensional one. In
principle all functionals may be either local or nonlocal.

            \setcounter{page}{1}
            \section{Basics of general covariant BV formalism}

The basic object in \fa quantization is the nilpotent fermionic second
order differential  operator
\be
&&\Delta\equiv\half(-1)^{\ep_{A}}\rho^{-1}\d_{A}\rho E^{AB}\d_{B},
\e{1}
where $\d_A$ are derivatives with respect to  local coordinates 
$\Ga^A$, $A=1,\ldots,2N$, on an antisymplectic manifold $\cM$. Their 
Grassmann parities are $\ep(\Ga^A)\equiv\ep_{A}\in\{0,1\}$. ($\Ga^A$ 
are  generalized fields and antifields.) $\rho(\Ga)$ is a measure
density and $E^{AB}$ an odd metric tensor with the properties: 
$E^{AB}=E^{BA}(-1)^{\ep_{A}+\ep_{B}+\ep_{A}\ep_{B}}$ and 
$\ep(E^{AB})=\ep_{A}+\ep_{B}+1$.  
Another basic object in the \fa formalism is the antibracket given by
\be
&&(F,G)\equiv (-1)^{\ep_F}\Delta(FG)-(-1)^{\ep_F}(\Delta F)G-F\Delta
G=F\bd_AE^{AB}\d_BG
\e{2}
for arbitrary functions $F,G$ on $\cM$. It satisfies 
\be
&&\ep((F,G))=\ep_F+\ep_G+1,\quad(F,G)=-(G,F)(-1)^{(\ep_F+1)(\ep_G+1)},\nn\\
&&(F,GH)=(F,G)H+(-1)^{\ep_G(\ep_F+1)}G(F,H),\nn\\
&&((F,G),H)(-1)^{(\ep_F+1)(\ep_H+1)}+{\rm cycle}(F,G,H)\equiv0,\nn\\
&&\Delta(F,G)=(\Delta F,G)+(-1)^{(\ep_F+1)}(F,\Delta G).
\e{3}
The measure density $\rho$ satisfies also
\be
&&(-1)^{(\ep_A+\ep_C)}\rho^{-1}\d_A \rho E^{AB}\d_B\rho^{-1}\d_C\rho
E^{CD}=0.
\e{4}
All these properties follow from the nilpotency of the 
$\Delta$-operator \r{1}, \ie
$\Delta^2=0$.

On $\cM$ we assume the existence of a quantum master action $W$ satisfying
\be
&&\Delta e^{{i\over \hbar}W}=0\quad\Lra\quad\half(W,W)=i\hbar\Delta W.
\e{5}
In terms of $W$ there is another nilpotent second order differential
operator
$\sigma_W$ defined by \cite{MH}
\be
&&\sigma_WF\equiv {\hbar\over i}e^{-{i\over\hbar}W}\Delta
e^{{i\over\hbar}W}F=(W,F)-i\hbar\Delta F
\e{6}
which satisfies $\sigma_W^2=0$ and 
\be
&&\sigma_W(F,G)=(\sigma_W F,G)+(-1)^{\ep_F+1}(F,\sigma_W G),\nn\\
&&\sigma_W FG=(\sigma_W F)G+(-1)^{\ep_F}F(\sigma_W
G)-i\hbar(-1)^{\ep_F}(F,G).
\e{7}
Given two solutions $W$ and $X$ of the master equation \r{5}, one has
\be
&&[\sigma_W,\sigma_X]F=-\half\L((-W+X,-W+X),F\R).
\e{71}
The path integral
in this generalized BV formalism is given by
\be
&&Z=\int\exp{\L\{{i\over\hbar}[W+X]\R\}}\rho[d\Ga][d\eta],
\e{8}
where $W$ is the above quantum master action and $X$ a
 hyper gauge-fixing master
action which also satisfies the quantum master equation 
\r{5}. $\eta^a$ are second
level Lagrange multipliers \cite{BT94-1,BT94-2} with no 
corresponding antifields. This
means that
$\cM$ is   viewed as containing first level 
Lagrange multipliers $\la^a$ and their
antifields
$\la^*_a$ ($\{\Ga^A\}=\{\Ga_0^A,\la^a, \la^*_a\}$) with the 
Grassmann parities 
$\ep(\eta^a)=\ep(\la^*_a)=\ep(\la^a)+1$. The actions
$W$ and
$X$ have  then the form 
\be
&&W=W_0(\Ga_0)+\la^*_a\eta^a,\quad X=G_a(\Ga_0)\la^a+\ldots,
\e{9}
where $G_a$ are hyperconstraints that fixes the antifields in 
$\{\Ga_0^A\}$. Also
$W_0$ satisfies the master equation \r{5}. 
Under these conditions one  may show that
the path integral
\r{8} is independent of the precise form of the hypergauge 
conditions  $G_a$
($X$-independence)
\cite{Trip}.

If we introduce some constraints $\Theta^\al=0$, 
$\al=1,\ldots,2n<2N$, on $\cM$
such that $E^{\al\beta}\equiv (\Theta^\al,\Theta^\beta)$ is 
invertible, then we may
define a ``Dirac" antibracket by the expression \cite{BT93}
\be
&&(F,G)_{(\cD)}\equiv(F,G)-(F,\Theta^\al)E_{\al\beta}
(\Theta^\beta,G),
\e{10}
where $E_{\al\beta}$ is the invers to $E^{\al\beta}$. Obviously
\be
&&(F,G)_{(\cD)}=F\bd_AE^{AB}_{(\cD)}\d_BG, 
\quad E^{AB}_{(\cD)}\equiv
E^{AB}-E^{AC}(\d_C\Theta^\al) E_{\al\beta}   
(\Theta^\beta\bd_D)E^{DB}.
\e{11}
Since $(F,\Theta^\al)_{(\cD)}=0$ for any $F$ the 
metric $E^{AB}_{(\cD)}$ is
degenerate on $\cM$. However, even in terms of such a metric
there is a consistent
path integral and it is given by \cite{BT93}
\be
&&Z_{(\cD)}=\int\exp{\L\{{i\over\hbar}[W+X]\R\}}
\prod_\al\del(\Theta^\al)\rho_{(\cD)}[d\Ga][d\eta],
\e{12}
where now $W$ and $X$ satisfy the quantum master equations 
\r{5} with $\Delta$
replaced by
\be
&&\Delta_{(\cD)}\equiv \half(-1)^{\ep_{A}}
\rho_{(\cD)}^{-1}\d_{A}\rho_{(\cD)}
E_{(\cD)}^{AB}\d_{B}.
\e{13}
Thus, for antisymplectic second class constraints 
$\Theta^\al=0$ there is a
consistent formulation already. We shall now propose a 
consistent formulation  for
corresponding first class constraints.

\section{Field-antifield formalism with first class constraints}

Let us call $T_\al=0$
antisymplectic first 
 class constraints provided $T_\al$ satisfy
\be
&&(T_\al,T_\beta)=T_\ga  U_{\;\al\beta}^{\ga}.
\e{14}
In the presence of such constraints 
we propose the following path integral
\be
&Z_{T}&=\int\exp{\L\{{i\over\hbar}[W+X]\R\}}
\prod_\al\del(T_\al)\prod_\beta\del(\chi^\beta)\:{1\over{\rm
sdet}(\chi^\ga,T_\del)}\rho(\Ga)[d\Ga][d\eta]=\nn\\
&&=
\int\exp{\L\{{i\over\hbar}
[W+X+\baC_\al(\chi^\al,T_\beta)\cC^\beta+
T_\al\pi^\al+\xi_\al\chi^\al]\R\}}d\mu,
\nn
\\
&d\mu&\equiv\rho(\Ga)[d\Ga][d\eta][d\pi][d\xi][d\cC][d\baC],
\e{15}
where $\rho(\Ga)$ is a gauge independent
measure density, 
and where $\chi^\al=0$ are
gauge-fixing conditions to
$T_\al=0$, \ie
$(\chi^\ga,T_\del)$ is required to be invertible. 
The Grassmann parities of the
field variables in \r{15} are
\be
&&\ep(\pi^\al)=\ep(\cC^\al)=\ep(\baC_\al)=\ep_\al
\equiv\ep(T_\al),\quad\ep(\xi_\al)=\ep(\chi^\al)=\ep_\al+1.
\e{151}
  The
first class constraints
$T_\al$ are in addition to \r{14} required to satisfy
\be
&&\sigma_W T_\al=
T_\beta P^\beta_{\;\al}-i\hbar 
U_{\;\beta\al}^\beta(-1)^{\ep_\beta},\nn\\
&&\sigma_X T_\al=
T_\beta Q^\beta_{\;\al}-i\hbar 
U_{\;\beta\al}^\beta(-1)^{\ep_\beta},
\e{16}
which  also may be viewed as  conditions  on $W$
and $X$.
A  general representation of the path 
integral \r{15} is given in appendix A.

The path integral $Z_T$ is invariant under the supertransformation
\be
&&\del\Ga^A=(\Ga^A,T_\al)\cC^\al\mu,
\e{17}
where $\mu$ is an odd constant. It leads to
\be
&&\del
T_\al=T_\ga U_{\;\al\beta}^\ga\cC^\beta\mu,\quad\del\chi^\al
=(\chi^\al,T_\beta)\cC^\beta\mu,\nn\\
&&\del(\baC_\al(\chi^\al,
T_\beta)\cC^\beta)=\half\baC_\al(\chi^\al,T_\del)
 U_{\;\beta\ga}^\del\cC^\ga\cC^\beta(-1)^{\ep_\beta}+\half
T_\del\baC_\al(\chi^\al, 
U_{\;\beta\ga}^\del)\cC^\ga\cC^\beta(-1)^{\ep_\beta}
\e{18} 
and 
\be
&&\del(W+X)=2i\hbar(\Delta
T_\al- U_{\;\beta\al}^\beta(-1)^{\ep_\beta})\cC^\al\mu+
T_\beta(P^\beta_{\;\al}+Q^\beta_{\;\al})\cC^\al\mu
\e{19}
from
\r{16}.
Furthermore, it gives rise to the following Jacobian
\be
&&J=1+2(\Delta T_\al)\cC^\al\mu.
\e{20}
All these contributions from the transformation \r{17} in the 
integrand of $Z_T$ are
compensated by the transformations
\be
&\del\cC^\del=&-\half U_{\;\beta\ga}^\del\cC^\ga
\cC^\beta(-1)^{\ep_\beta}\mu,\nn\\
&\del\pi^\del=&-\half\baC_\al(\chi^\al,
U_{\;\beta\ga}^\del)\cC^\ga\cC^\beta(-1)^{\ep_\beta}\mu-\nn\\ &&-
U_{\;\al\beta}^\del\cC^\beta\pi^\al
(-1)^{\ep_\al}\mu-(P^\del_{\;\al}+Q^\del_{\;\al})\cC^\al\mu,\nn\\
&\del\baC_\al=&\mu\xi_\al,
\e{21}
together with the resulting contributions from the corresponding Jacobians
\be
&&\del\cC^\del{\bd\over\d\cC^\del}(-1)^{\ep_\del}=- 
U_{\;\beta\ga}^\beta\cC^\ga(-1)^{\ep_\beta}\mu,\quad
\del\pi^\del{\bd\over\d\pi^\del}(-1)^{\ep_\del}
=- U_{\;\beta\ga}^\beta\cC^\ga(-1)^{\ep_\beta}\mu.
\e{22}

The path integral $Z_T$ is also independent of the gauge-fixing 
functions $\chi^\al$.
To see this consider the shift
\be
&&\chi^\al\ra\chi^\al+\del\chi^\al
\e{23}
in $Z_T$. It is in fact exactly compensated 
by the transformation \r{17}  with the
choice
\be
&&\mu={i\over\hbar}\baC_\al\del\chi^\al,
\e{24}
since when compared to the previous 
transformation with $\mu$ constant this choice
gives rise to the following additional contribution to the Jacobian
\be
&\del
J&\equiv(\Ga^A,T_\al)\cC^\al{i\over\hbar}
\L(\baC_\beta\del\chi^\beta\bd_A\R)
(-1)^{\ep_A}+{i\over\hbar}(-1)^{\ep_\al}
{\d\over\d\baC_\al}(\baC_\beta\del\chi^\beta)\xi_\al=
\nn\\&&=
-{i\over\hbar}\baC_\beta(\del\chi^\beta,T_\al)
\cC^\al-{i\over\hbar}\xi_\al\del\chi^\al.
\e{25}

\section{Conversion and the Abelian case}
We shall now apply and verify the  general formulation above. We
consider then a generalized conversion of antisymplectic second class
constraints into corresponding first class ones. Within the
ordinary Hamiltonian  formalism the conversion mechanism has been
formulated in
general  terms in
\cite{BF,BFF} (see also
\cite{FS}). It has been applied to many models. 
One  interesting application is the
new approach to geometric quantization  in \cite{FL} which is mainly based
on
\cite{BFF}. In the following application  to the \fa formalism a new
ingredient
appears since we not only have  antibrackets, which
 corresponds to Poisson brackets,
but also the  nilpotent differential $\Delta$-operator.

Consider  the second class constraints
$\Theta^\al=0$, $\al=1,\ldots,2n<2N$, on $\cM$ which by definition are
such that $E^{\al\beta}\equiv 
(\Theta^\al,\Theta^\beta)$ is invertible. We  now
convert these constraints into 
abelian first class constraints by extending the
original antisymplectic manifold $\cM$. Introduce therefore 
the additional \fa coordinates 
$\Phi^\al$ with the Grassmann parities
$\ep(\Phi^\al)=\ep(\Theta^\al)=\ep_\al$. 
On the resulting extended manifold,
$\cM\ex\equiv\cM\oplus\{\Phi^\al\}$, we define then an extended 
antibracket with the
extended metric 
\be
&&(\Ga^A,\Ga^B)\ex=E^{AB},
\quad(\Ga^A,\Phi^\al)\ex=0,\quad(\Phi^\al,\Phi^\beta)\ex=\omega^{\al\beta},
\e{26}
where $\omega^{\al\beta}$ is an odd invertible constant matrix.
On $\cM\ex$ we may then define first class constraints $T^\al$ satisfying
\be
&&(T^\al,T^\beta)\ex=0, \quad T^\al|_{\Phi=0}=\Theta^\al.
\e{27}
These functions may be
constructed perturbatively with the ansatz
\be
&&T^\al(\Ga,\Phi)=\Theta^\al(\Ga)+\sum_{n=1}^\infty{1\over
n!}\Phi^{\beta_n}\cdots\Phi^{\beta_1}
X_{\beta_1\cdots\beta_n}^{\qquad\al}(\Ga).
\e{28}
We may furthermore construct gauge
invariant functions
$\bar{G}(\Ga,\Phi)$ to any function $G(\Ga)$ by the conditions
\be
&&(\bar{G},T^\al)\ex=0,\quad\bar{G}|_{\Phi=0}=G.
\e{29}
Also these conditions may be solved 
perturbatively with an ansatz of the form
\be
&&\bar{G}(\Ga,\Phi)=G(\Ga)+\sum_{n=1}^\infty{1\over
n!}\Phi^{\beta_n}\cdots\Phi^{\beta_1}Y_{\beta_1\cdots\beta_n}(\Ga).
\e{30}
In appendix B it is shown that
\be
&&(\bar{F},\bar{G})\ex|_{\Phi=0}=(F,G)_{(\cD)},
\e{31}
where the right-hand side is the Dirac antibracket \r{10}.
One may  also show that (see below)
\be
&&\Delta\ex\bar{G}|_{\Phi=0}=\Delta_{(\cD)}G
\e{32}
provided
\be
&&\Delta\ex T^\al=0.
\e{36}
 $\Delta\ex$ is the corresponding 
$\Delta$-operator to \r{1} on the extended
manifold
$\cM\ex$ with the metric \r{26} and with
 a measure density $\rho(\Ga,\Phi)$
satisfying \r{36}. To show
\r{31} it is sufficient to solve \r{28} and \r{30} 
up to the first order in $\Phi$
as is shown in  appendix B, while \r{32} requires 
a solution up to second order.

The
 gauge independent path integral \r{15} is here given by 
\be
&&Z_{T}=\int\exp{\L\{{i\over\hbar}[{W}\ex+{X}\ex]\R\}}
\prod_\al\del(T^\al)\prod_\beta\del(\chi_\beta)\:{1\over{\rm
sdet}(\chi_\ga,T^\del)}\rho(\Ga,\Phi)[d\Ga][d\Phi][d\eta],\nn\\
\e{33}
where ${W}\ex$ and ${X}\ex$ satisfy the quantum master equations
\be
&&\Delta\ex e^{{i\over\hbar}{W}\ex}=0,\quad 
\Delta\ex e^{{i\over\hbar}{X}\ex}=0.
\e{34}
According to \r{16} we have additional conditions like (see appendix A)
\be
&&({W}\ex,T^\al)\ex=i\hbar\Delta\ex 
T^\al,\quad({X}\ex,T^\al)\ex=i\hbar\Delta\ex T^\al.
\e{35}
They are easy to solve for measure densities satisfying \r{36}
since these conditions then reduce to
\be
&&(W\ex,T^\al)\ex=(X\ex,T^\al)\ex=0.
\e{37}
This implies that ${W}\ex$ and ${X}\ex$ in this case 
are gauge invariant extensions
of 
${W}$ and ${X}$ defined by
\be
&&{W(\Ga)}\equiv{W}\ex|_{\Phi=0},\quad {X(\Ga)}\equiv{X}\ex|_{\Phi=0},
\e{38}
which means that ${W}\ex=\bar{W}$ and ${X}\ex=\bar{X}$.
It follows now that the path integral \r{33} in
 \eg the gauge $\chi_\al=\omega_{\al\beta}\Phi^\beta$  reduces to
\be
&&Z_{T}=\int\exp{\L\{{i\over\hbar}[W+X]\R\}}
\prod_\al\del(\Theta^\al)\:{1\over{\rm
sdet}(X_\ga^{\;\del})}\rho(\Ga,0)[d\Ga][d\eta],
\e{39}
which when compared to the second class expression 
$Z_{(\cD)}$ in \r{12} requires the
boundary condition 
\be
&&\rho(\Ga,0)=\rho_{(\cD)}(\Ga)\:{\rm sdet}(X_\ga^{\;\del}),
\e{40}
where $X_\ga^{\;\del}$ is the first order coefficient in 
the expansion \r{28}.
That $W$ and $X$ satisfy the appropriate master equations 
follows from \r{31} and
\r{32}.

Another equivalent but more explicit and transparent way 
to derive the equivalence
between
\r{12} and \r{33} is to first construct gauge invariant coordinates
$\baGa^A(\Ga,\Phi)$ defined by
\be
&&(\baGa^A,T^\al)\ex=0,\quad \baGa^A|_{\Phi=0}=\Ga^A,
\e{41}
which also may be solved by a 
perturbative ansatz like \r{30}. Then we have
$\bar{G}=G(\baGa)$ for any gauge invariant function and
\be
&&(\baGa^A,\baGa^B)\ex|_{\Phi=0}=E^{AB}_{(\cD)}.
\e{42}
Thus, the gauge invariant functions lives on 
the submanifold of $\cM\ex$ spanned by
$\baGa^A$. The same is true for the 
$\Delta$-operator as will be shown below.

It is  convenient to change coordinates on
 $\cM\ex$ from $\{\Ga,\Phi\}$ to
$\{\baGa,\Phi\}$. In terms of these coordinates we have
\be
&&(\bar{F},\bar{G})\ex=F(\baGa)\bbad_A
\bar{E}^{AB}\bad_BG(\baGa),\nn\\
&&\bar{E}^{AB}\equiv(\baGa^A,\baGa^B)\ex,
\e{43}
where $\bad_A$ are derivatives with respect to $\baGa^A$.
Furthermore, we have
\be
&&{\Delta}\ex=\half(-1)^{\ep_A}\bar{\rho}^{-1}
\bad_A\bar{\rho}\bar{E}^{AB}\bad_B+
\half(-1)^{\ep_A}\bar{\rho}^{-1}\bad_A\bar{\rho}
\bar{E}^{A\beta}\bad_\beta+\nn\\
&&+\half(-1)^{\ep_\al}\bar{\rho}^{-1}\bad_\al
\bar{\rho}\bar{E}^{\al B}\bad_B+
\half(-1)^{\ep_\al}\bar{\rho}^{-1}\bad_\al\bar{\rho}
\bar{E}^{\al\beta}\bad_\beta,
\e{44}
where  $\bad_\al$  are derivatives with respect to 
$\Phi^\al$ while keeping $\baGa^A$
fixed, and where $\bar{\rho}$ is related to 
$\rho$ through the formula
\be
&&\bar{\rho}(\baGa,\Phi)\:{\rm sdet}(\d_A\baGa^B)=\rho(\Ga,\Phi).
\e{45}
The ``bar"-metric in \r{44}  is given in \r{43} and
\be
&&\bar{E}^{A\beta}\equiv(\baGa^A,\Phi^\beta)\ex=
\baGa^A\bd_\al\omega^{\al\beta},\quad
\bar{E}^{\al B}\equiv(\Phi^\al,\baGa^B)\ex=
\omega^{\al\beta}\d_\beta\baGa^B,\nn\\
&&\bar{E}^{\al \beta}\equiv(\Phi^\al,
\Phi^\beta)\ex=\omega^{\al\beta}.
\e{46}
Since the Jacobi identities yield
\be
&&(\bar{E}^{AB},T^\al)\ex=0,
\e{48}
we have also
\be
&&\bar{E}^{AB}={E}^{AB}_{(\cD)}(\baGa)
\e{49}
due to \r{42}.
The $\Delta$-operator expression \r{44} may be decomposed as follows
\be
&&{\Delta}\ex=\bar{\Delta}+K^\al\bad_\al,\nn\\
&&\bar{\Delta}=
\half(-1)^{\ep_A}\bar{\rho}^{-1}\bad_A
\bar{\rho}\bar{E}^{AB}\bad_B+
\half(-1)^{\ep_\al}\bar{\rho}^{-1}(\bad_\al
\bar{\rho}\bar{E}^{\al
B})\bad_B.
\e{47}
Obviously ${\Delta}\ex\bar{F}=\bar{\Delta}\,\bar{F}$ 
for any gauge invariant
function $\bar{F}$.
Furthermore, since the condition
${\Delta}\ex
T^\al\equiv{\Delta}\ex\Theta^\al(\baGa)=
\bar{\Delta}\Theta^\al(\baGa)=0$ implies that
\be
&&\half(-1)^{\ep_\al}\bar{\rho}^{-1}(\bad_\al\bar{\rho}\bar{E}^{\al
B})=\half(-1)^{\ep_A}F_A\bar{E}^{AB}
\e{50}
for any function $F_A$, we have 
\be
&&\bar{\Delta}=
\half(-1)^{\ep_A}(\bad_A+F_A+(\bad_A\ln\bar{\rho}))\bar{E}^{AB}\bad_B.
\e{51}
Therefore, if we restrict $F_A$ such that $F_A+(\bad_A\ln\bar{\rho})$ 
only depends on
$\baGa^A$ then also $\bar{\Delta}$ only depends on
$\baGa^A$ due to \r{49}. The nilpotency of 
$\Delta\ex$ requires then that
$\bar{\Delta}$ is
 nilpotent which in turn implies \cite{BT94-2}
\be
&&F_A+(\bad_A\ln\bar{\rho})=\bad_A\ln\tilde{\rho}(\baGa).
\e{52}
This result may equivalently be expressed as follows: 
In order for the measure density
$\rho$ to satisfy
\r{36} it should be such that
\be
&&\rho[d\Ga][d\Phi]=\bar{\rho}[d\bar{\Ga}][d\Phi]=
\tilde{\rho}(\bar{\Ga})[d\bar{\Ga}][d
T^*]
\e{521}
for any measure density $\tilde{\rho}(\bar{\Ga})$ where
$T^*_\al$ satisfies
\be
&&(T^\al,T^*_\beta)\ex=\del_\beta^\al,
\quad(T^*_\al,T^*_\beta)\ex=0,\quad\ep(T^*_\al)=
\ep_\al+1.
\e{522}
We assert that there is a solution of the form
\be
&&T^*_\al(\Ga,\Phi)=\Theta^*_\al(\Ga)+\sum_{n=1}^\infty{1\over
n!}\Phi^{\beta_n}\cdots
\Phi^{\beta_1}X^*_{\beta_1\cdots\beta_n\al}(\Ga).
\e{523}
where also the function $\Theta^*_\al(\Ga)$
is to be determined (see appendix C).
Obviously
$\Delta\ex\bar{G}|_{\Phi=0}=
\bar{\Delta}\,\bar{G}|_{\Phi=0}=\Delta_{(\cD)}G$ in
agreement with the assertion \r{32}, provided
\be
&&\tilde{\rho}(\baGa)=\rho_{(\cD)}(\baGa),
\e{53}
where $\rho_{(\cD)}$ is the Dirac measure density in \r{12} and \r{13}.
Thus, the $\Delta$-operator 
$\bar{\Delta}$ is  just a gauge invariant
extension of
$\Delta_{(\cD)}$ on
$\cM\ex$.
It should also be mentioned that the 
transformation \r{521} with the identification
\r{53} is consistent with the boundary 
condition \r{40} due to the relation
\be
&&(T^*_\al,\Phi^\beta)\ex|_{\Phi=0,\Theta=0}
=-\half(X^{-1})_\al^{\;\beta},
\e{54}
which follows from \r{522} and \r{523} 
to lowest order in $\Phi^\al$ (see
formula \r{c9} in appendix C).
\vspace{1cm}

\newpage
{\bf Acknowledgements}\\

I.A.B. thanks Klaus Bering and Poul Damgaard
 for stimulating discussions at an early
stage of this work. I.A.B. would also like to thank
 Lars Brink for very
warm hospitality at the Institute of Theoretical
Physics, Chalmers and G\"{o}teborg
University. The work is partially supported by grant INTAS-RFBR 95-0829. 
The work of I.A.B. is also 
supported by grants 
INTAS  93-2058, INTAS 93-0633, RFBR 96-01-00482, 
RFBR 96-02-17314, and NorFa 97.40.002-O.\\ \\

\def\theequation{\thesection.\arabic{equation}}
\setcounter{section}{1}
\setcounter{equation}{0}
\renewcommand{\thesection}{\Alph{section}}

    \noindent
    {\Large{\bf{Appendix A}}}\\ \\
 {\bf Invariant formulation of the path integral $Z_T$}\\ \\
Let us extend the antisymplectic manifold $\cM$ 
in section 2 by three sets of \fa
pairs:
$\{\cC^*_\al,\cC^\al;
\pi_\al^*,\pi^\al;\baC^*_\al,\baC^{\al}\}$. Their
Grassmann parities are
\be
&&\ep{(\cC^\al)}=\ep{(\pi^\al)}=
\ep{(\baC^{*}_\al)}=\ep_{\al}\equiv\ep{(T^\al)},
\quad\ep{(\cC^*_\al)}=\ep{(\pi_\al^*)}=\ep{(\baC^\al)}=\ep_{\al}+1.
\e{a1}
On this
extended manifold, $\tilde{\cM}$, we define the bosonic charge 
\be
&&\Omega=\L(
T_\al\cC^\al-\half\cC^*_\al
 U^\al_{\;\beta\ga}\cC^\ga\cC^\beta(-1)^{\ep_\beta}
+\ldots\R)-\pi_\al^*\baC^{\al},
\e{a2}
where $T_\al$ satisfies the algebra \r{14} 
and where the dots indicate terms
independent of
$\pi^\al$ and
$\baC^*_\al$ such that
$\Omega$ satisfies
\be
&&(\Omega,\Omega)=0,
\e{a3}
where the antibracket from now on is the 
extended one on  $\tilde{\cM}$. An invariant 
path integral
 may then be written as
\be
&&Z_T=\int \exp{\L\{{i\over\hbar}[\tilde{W}+
\tilde{X}+(\Psi,\Omega)]\R\}}\tilde{\rho}
[d\Ga][d\eta][d\cC][d\cC^*][d\pi][d\pi^*][d\baC][d\baC^*],
\e{a4}
where $\tilde{W}$ and $\tilde{X}$ 
satisfy the extended master equations
\be
&&\tilde{\Delta}e^{{i\over\hbar}\tilde{W}}=0,\quad\tilde{\Delta}
e^{{i\over\hbar}\tilde{X}}=0,
\e{a5}
where in turn $\tilde{\Delta}$ is 
the nilpotent $\Delta$-operator \r{1} extended to
$\tilde{\cM}$ with $\tilde{\rho}$. The objects
$\Omega$,
$\tilde{W}$ and
$\tilde{X}$ are in addition required to satisfy
\be
&&\sigma_{\tilde{W}}\Omega=\sigma_{\tilde{X}}\Omega=0.
\e{a7}
The gauge-fixing charge $\Psi$ is odd and is of the form
\be
&&\Psi=\baC^*_\al\chi^\al-\cC^*_\al\pi^\al.
\e{a8} 
The solutions of \r{a5} are 
\be
&&\tilde{W}=W(\Ga)-\cC^*_\al
P^\al_{\;\beta}\cC^\beta+\ldots,\quad\tilde{X}=X(\Ga)-\cC^*_\al
Q^\al_{\;\beta}\cC^\beta+\ldots,
\e{a6}
The equations for $W$ and $X$ reduce to \r{5} if and only if
 $P^\al_{\;\al}=Q^\al_{\;\al}=0$. Otherwise $W$ and $X$ generalize 
to satisfy the modified equations with supertrace
``anomalies" in their right-hand sides (see appendix B). 
The formulation given in section 2
corresponds therefore to the first case. Notice that 
eq.\r{16} is obtained from 
\r{a7} when \r{a6} is inserted, and that with the choice \r{a8} 
the path integral \r{a4} reduces to
\r{15} after the identifications 
$\pi^*_\al\equiv\xi_\al$,
$\baC_\al^*\equiv\baC_\al$, and provided
$\tilde{\rho}={\rho}(\Ga)$ and 
$\chi^\al$ only depends on $\Ga^A$.

The path integral \r{a4} is invariant 
under the following transformation
\be
&&\del\tilde{\Ga}^\cA=(\tilde{\Ga}^\cA,\Omega)\mu,
\e{a9}
where
$\tilde{\Ga}^{\cA}\equiv\{\Ga^A;\cC^*_\al,\cC^\al;
\pi_\al^*,\pi^\al;\baC^{*}_{\al},\baC^\al\}\in\tilde{\cM}$ and
where $\mu$ is an odd constant. 
The contribution to the Jacobian 
\be
&&J-1=2(\tilde{\Delta}\Omega)\mu
\e{a10}
 is compensated by corresponding 
terms from \r{a7}. The path integral \r{a4} is
also independent of $\Psi$ since 
$\Psi\ra\Psi+\del\Psi$ is compensated by the
additional contribution to the Jacobian 
from the transformation \r{a9} with the
choice
$\mu=i\del\Psi/\hbar$. Furthermore, 
\r{a4} is independent of $\tilde{X}$ which
contains the hypergauge-fixing. 
To see this consider the transformation
\be
&&\del\tilde{\Ga}^\cA=(\tilde{\Ga}^\cA,-
\tilde{W}+\tilde{X})\xi+{\hbar\over
i}(\tilde{\Ga}^\cA,\xi),
\e{a11}
where $\xi$ is an odd infinitesimal 
function satisfying the condition
\be
&&(\xi,\Omega)=0.
\e{a12}
The Jacobian of \r{a11} is
\be
&&J-1=2(-\tilde{\Delta}\tilde{W}+
\tilde{\Delta}\tilde{X})\xi+2{\hbar\over
i}\tilde{\Delta}\xi+(-\tilde{W}+\tilde{X},\xi).
\e{a13}
Summing up the total contribution from 
\r{a11} in \r{a4} one finds after use of the
master equations \r{a5} that what remains 
may be viewed as the following
transformations
\be
&&\del\Psi=(\Psi,-\tilde{W}+\tilde{X})\xi+{\hbar\over i}(\Psi,\xi),\quad
\del\tilde{X}=2{\hbar\over i}\sigma_{\tilde{X}}\xi.
\e{a14}
However, since $Z_T$ is independent of $\Psi$
 as was shown above only $\del\tilde{X}$
remains. Notice that $(\del\tilde{X},\Omega)=0$
 in consistency with \r{a7}.

One may also notice that \r{a4} is invariant under
general anticanonical transformations of the form
\be
&&\del\tilde{\Ga}^\cA=(\tilde{\Ga}^\cA,G)
\e{a15}
for any odd infinitesimal function $G$ provided 
$\tilde{W}$ and $\tilde{X}$ transform
according to
\be
&&\del\tilde{W}=\sigma_{\tilde{W}}G,
\quad\del\tilde{X}=\sigma_{\tilde{X}}G.
\e{a16}
This may be used to demonstrate the 
existence of the above formulation. Consider the
abelian case
\be
&&\Omega_{\rm Abel}=
t_\al\cC^\al-\pi_\al^*\baC^{\al}, 
\quad (\Omega_{\rm Abel},\Omega_{\rm Abel})=0.
\e{a17}
Let the master actions here satisfy $\tilde{W}=
{W}$ and $\tilde{X}={W}$. (This
implies that
\r{a7} reduces to
\r{35}.) We define then  $\Omega$, $\tilde{W}$,
 $\tilde{X}$ in terms of a finite
transformation of $\Omega_{\rm Abel}$, ${W}$,
 ${X}$ of the form \r{a15}-\r{a16}, \ie
\be
&&\Omega\equiv\exp{(G,\cdot)}\Omega_{\rm Abel},\quad
e^{{i\over\hbar}\tilde{W}}=
e^{[\tilde{\Delta},G]}e^{{i\over\hbar}{W}},\quad
e^{{i\over\hbar}\tilde{X}}=
e^{[\tilde{\Delta},G]}e^{{i\over\hbar}{X}}.
\e{a18}
Obviously $(\Omega,\Omega)=0$, and if
 $W$, $X$ satisfies \r{a7} with $\Omega_{\rm
Abel}$ then $\tilde{W}$, $\tilde{X}$ satisfy 
\r{a7} with $\Omega$. In this manner
the non-abelian case is obtained by 
choosing the anticanonical generator $G$ in the
following form 
\be
&&G=\cC^*_\al \Lambda^\al_{\;\beta}\cC^\beta,
\e{a19}
where the matrix $e^\Lambda$ changes 
effectively the constraint basis.\\ \\

\setcounter{equation}{0}
\setcounter{section}{2}
\noindent
    {\Large{\bf{Appendix B}}}\\ \\
 {\bf The path integral \r{15} with anomalous master actions}\\ \\
Consider the path integral \r{15} in section 2 
where $T_\al$ satisfies the algebra \r{14} and the
properties \r{16}. According to appendix A the
 master actions $W$ and $X$ do no longer
satisfy the master equation \r{5} if 
$P^\al_{\;\al}\neq0$ and $Q^\al_{\;\al}\neq0$.
The appropriate generalized master equations in the latter case are
\be
&&(i\hbar\Delta +P^\al_{\;\al}) e^{{i\over \hbar}W}=0\quad\Lra\quad
\half(W,W)=i\hbar\Delta W-i\hbar P^\al_{\;\al},\nn\\
&&(i\hbar\Delta +Q^\al_{\;\al}) e^{{i\over \hbar}X}=0\quad\Lra\quad
\half(X,X)=i\hbar\Delta X-i\hbar Q^\al_{\;\al}.
\e{b1}
Since a consistent theory requires $\Delta$ to be 
nilpotent we have from \r{b1} the
consistency conditions
\be
&&\Delta(P^\al_{\;\al} e^{{i\over \hbar}W})=0\quad\Lra\quad
\sigma_W P^\al_{\;\al}=0,\nn\\
&&\Delta (Q^\al_{\;\al} e^{{i\over \hbar}X})=0\quad\Lra\quad
\sigma_X Q^\al_{\;\al}=0.
\e{b2}

The proof of the invariance under \r{17} as well
 as the independence of the gauge
fixing function $\chi^\al$ given in \r{23}-\r{25}
 are still valid in this generalized
case. We may also prove the independence of the 
gauge fixing action $X$ following the
argument of appendix A in a reduced form. We perform then the
following change of integration variables in the path integral \r{15}:
\be
&\del\Gamma^A=&(\Gamma^A,-W+X)\mu+{\hbar\over
i}(\Gamma^A,\mu),\quad\del\cC^\al=
(P^\al_{\;\beta}-Q^\al_{\;\beta})\,\cC^\beta
\mu+{\hbar\over i}R^\al_{\;\beta}\,\cC^\beta,\nn\\
&\del\pi^\al=&(P^\al_{\;\beta}-Q^\al_{\;\beta})\,\pi^\beta \mu+{\hbar\over
i}R^\al_{\;\beta}\,\pi^\beta+{\hbar\over
i}\baC_\beta\,(\chi^\beta,R^\al_{\;\ga})\,\cC^\ga+\nn\\&&+\baC_\beta\,
(\chi^\beta,P^\al_{\;\ga}
-Q^\al_{\;\ga})\,\cC^\ga
\mu,
\e{b3}
where $\mu$ is an odd function which satisfies the condition
\be
&&(\mu, T_\al)=T_\beta R^\beta_{\;\al},
\e{b4}
which in turn determines $R^\beta_{\;\al}$. 
The change of integration variables \r{b3}
in
\r{15} results in the following change in $X$:
\be
&&\del X={2\hbar\over i}\left[\sigma_X \mu-i\hbar
 R^\al_{\;\al}(-1)^{\ep_\al}\right],
\e{b5}
together with the following variation in $\chi^\al$:
\be
&&\del\chi^\al=(\chi^\al,-W+X)\mu+{\hbar\over i}(\chi^\al, \mu),
\e{b6} 
which is inessential as the previous proof of 
$\chi$-independence remains valid.

The change $\del X$ in \r{b5} is the most general one
 compatible with the allowed
variation
\be
&&\del Q^\al_{\;\al}={2\hbar\over i}\left[\sigma_X
R^\al_{\;\al}(-1)^{\ep_\al}+(Q^\al_{\;\al},
\mu)\right]
\e{b7}
in the modified equation for $X$ given in \r{b1}.
 This variation is the trace of the
matrix variation
\be
&&\del Q^\al_{\;\beta}={2\hbar\over i}\left[\sigma_X
R^\al_{\;\beta}(-1)^{\ep_\al}+(Q^\al_{\;\beta},
\mu)+Q^\al_{\;\ga} R^\ga_{\;\beta}-R^\al_{\;\ga} Q^\ga_{\;\beta}+i\hbar
S^{\al\ga}_{\ga\beta}\right],
\e{b8}
where in turn $S^{\al\ga}_{\ga\beta}$ is determined by the relation
\be
&&(R^\ga_{\;\al}, T_\beta)-(R^\ga_{\;\beta},
T_\al)(-1)^{(\ep_\al+1)(\ep_\beta+1)}+(-1)^{\ep_\ga}(\mu,
U^\ga_{\;\al\beta})-\nn\\
&&-R^\ga_{\;\del} U^\del_{\;\al\beta}+U^\ga_{\;\al\del}
R^\del_{\;\beta}-U^\ga_{\;\beta\del}
R^\del_{\;\al}(-1)^{(\ep_\al+1)(\ep_\beta+1)}+T_\del
S^{\del\ga}_{\al\beta}(-1)^{\ep_\ga}=0,
\e{b9}
which is a compatibility condition to \r{b4}. 
In terms of $\del Q^\al_{\;\beta}$ we
have
\be
&&(\del X,T_\al)=T_\beta \,\del Q^\beta_{\;\al}.
\e{b10}
The main part of these formulas may be derived from 
\r{a11}-\r{a14} with the ansatz
\be
&&\xi=\mu+\cC^*_\al R^\al_{\;\beta}\cC^\beta+{1\over
4}\cC^*_\al\cC^*_\beta S^{\beta\al}_{\del\ga}
\cC^\ga\cC^\del(-1)^{\ep_\del}+\ldots.
\e{b11}

\setcounter{equation}{0}
\setcounter{section}{3}
    \noindent
    {\Large{\bf{Appendix C}}}\\ \\
 {\bf Proof of formula \r{31}}\\ \\
Inserting the ansatz \r{28} into \r{27} one 
finds to the zeroth order in $\Phi^\al$ the
condition
\be
&&(\Theta^\al,\Theta^\beta)\equiv
E^{\al\beta}=-(-1)^{\ep_\ga(1+\ep_\al)}
X_\ga^{\;\al}\omega^{\ga\rho}X_\rho^{\;\beta}
\e{c1}
for the first order coefficients 
$X_\beta^{\;\al}(\Ga)$ in \r{28}. This implies that
\be
&&E_{\al\beta}=-(-1)^{\ep_\beta(1+\ep_\ga)}(X^{-1})_\al^{\;\rho}
\omega_{\rho\ga}(X^{-1})_\beta^{\;\ga},
\e{c2}
where $(X^{-1})_\al^{\;\rho}$ 
is the inverse to $X_\beta^{\;\al}$ in the sense
\be
&&(X^{-1})_\al^{\;\ga}X_\ga^{\;\beta}=
X_\al^{\;\ga}(X^{-1})_\ga^{\;\beta}=\del_\al^\beta.
\e{c3}
For the gauge invariant functions 
$\bar{F}$ and $\bar{G}$ in \r{31} we have to the
first order in
$\Phi$
\be
&&\bar{F}=F(\Ga)+\Phi^\al Y_\al(F)+O(\Phi^2),\quad\bar{G}=
G(\Ga)+\Phi^\al Y_\al(G)+O(\Phi^2),
\e{c4}
where
\be
&Y_\al(G)&=-(-1)^{\ep_\al(1+\ep_G)}(G,
\Theta^\ga)(X^{-1})_\ga^{\;\beta}\omega_{\beta\al}
=\nn\\
&&=-(-1)^{\ep_\ga(1+\ep_\beta)}
\omega_{\al\beta}(X^{-1})_\ga^{\;\beta}(\Theta^\ga,G).
\e{c5}
These expressions and \r{c2} imply now
\be
&&(\bar{F},\bar{G})\ex|_{\Phi=0}=(F,G)+(-1)^{\ep_\al(1+\ep_F)}
Y_\al(F)\omega^{\al\beta}Y_\beta(G)=\nn\\
&&=(F,G)-(F,\Theta^\al)E_{\al\beta}
(\Theta^\beta,G)\equiv(F,G)_{(\cD)}.
\e{c6}\\ \\
 \noindent
 {\bf Equation for $\Theta^*_\al$ in \r{523}}\\ \\
In parallel   to \r{28}, if one inserts 
the ansatz \r{523} into eq.\r{522} one gets
to the zeroth order in $\Phi^\al$ the equations
\be
&&(\Theta^\al,\Theta^*_\beta)+(-1)^{\ep_\mu(1+\ep_\al)}
X^{\;\al}_\mu\omega^{\mu\nu}X^*_{\nu\beta}=\del^\al_\beta,
\e{c7}
\be
&&(\Theta^*_\al,
\Theta^*_\beta)+(-1)^{\ep_\mu\ep_\al}
X^*_{\mu\al}\omega^{\mu\nu}X^*_{\nu\beta}=0.
\e{c8}
Solving \r{c7} for $X^*_{\al\beta}$ one finds
\be
&&X^*_{\al\beta}= - \L(\del_\beta^\ga+
(\Theta^*_\beta, \Theta^\ga)   \R)
(X^{-1})_\ga^{\;\del}
\omega_{\del\al}(-1)^{\ep_\al\ep_\beta}, 
\e{c9}
which when inserted into \r{c8} results in 
the following equation for $\Theta^*_\al$:
\be
&&(\Theta^*_\al,\Theta^*_\beta)_{(\cD)}=-E_{\al\beta}-
(\Theta^*_\al,\Theta^\ga)E_{\ga\beta}+
E_{\al\ga}(\Theta^\ga,\Theta^*_\beta).
\e{c10}
The solution is of the form
\be
&&\Theta^*_\al=\half\Theta^\ga E_{\ga\al}+O(\Theta^2).
\e{c11}

To confirm the existence of $T^*_\al$ satisfying \r{522}
 we notice that for abelian second class
constraints $\theta^\al(\Ga)=0$ satisfying 
$(\theta^\al,\theta^\beta)=
-\omega^{\al\beta}$ we have the following explicit
conversion formula for corresponding abelian 
first class constraints $t^\al$ and $t^*_\al$:
\be
&&t^\al=\theta^\al+\Phi^\al,\quad
t^*_\al=-\half(\theta^\ga-\Phi^\ga)\omega_{\ga\al}
\e{c12}
The general abelian functions $T^\al$ and $T^*_\al$ are then obtained
from $t^\al$ and $t^*_\al$ through the formula
\be
&&T^\al=\exp{(G,\cdot)\ex}\,t^\al,
\quad T^*_\al=\exp{(G,\cdot)\ex}\,t^*_\al
\e{c13}
where $G(\Ga,\Phi)$ is an odd function.

\end{document}